# De-amortized Cuckoo Hashing: Provable Worst-Case Performance and Experimental Results

Yuriy Arbitman*       Moni Naor†       Gil Segev‡

May 28, 2018

**Abstract**

Cuckoo hashing is a highly practical dynamic dictionary: it provides amortized constant insertion time, worst case constant deletion time and lookup time, and good memory utilization. However, with a noticeable probability during the insertion of $n$ elements some insertion requires $\Omega(\log n)$ time. Whereas such an amortized guarantee may be suitable for some applications, in other applications (such as high-performance routing) this is highly undesirable.

Recently, Kirsch and Mitzenmacher (Allerton '07) proposed a de-amortization of cuckoo hashing using various queueing techniques that preserve its attractive properties. Kirsch and Mitzenmacher demonstrated a significant improvement to the worst case performance of cuckoo hashing via experimental results, but they left open the problem of constructing a scheme with provable properties.

In this work we follow Kirsch and Mitzenmacher and present a de-amortization of cuckoo hashing that *provably* guarantees constant worst case operations. Specifically, for any sequence of polynomially many operations, with overwhelming probability over the randomness of the initialization phase, each operation is performed in constant time. Our theoretical analysis and experimental results indicate that the scheme is highly efficient, and provides a practical alternative to the only other known dynamic dictionary with such worst case guarantees, due to Dietzfelbinger and Meyer auf der Heide (ICALP '90).

*Email: yuriy.arbitman@gmail.com.

†Incumbent of the Judith Kleeman Professorial Chair, Department of Computer Science and Applied Mathematics, Weizmann Institute of Science, Rehovot 76100, Israel. Email: moni.naor@weizmann.ac.il. Research supported in part by a grant from the Israel Science Foundation.

‡Department of Computer Science and Applied Mathematics, Weizmann Institute of Science, Rehovot 76100, Israel. Email: gil.segev@weizmann.ac.il. Research supported by the Adams Fellowship Program of the Israel Academy of Sciences and Humanities, and by a grant from the Israel Science Foundation.

## 1 Introduction

A dynamic dictionary is a fundamental data structure used for maintaining a set of elements under insertions and deletions, while supporting membership queries. The performance of a dynamic dictionary is measured mainly by its update time, lookup time, and memory utilization. Extensive research has been devoted over the years for studying dynamic dictionaries both on the theoretical side by exploring upper and lower bounds on the performance guarantees, and on the practical side by designing efficient dynamic dictionaries that are suitable for real-world applications.

The most efficient dictionaries, in theory and in practice, are based on various forms of hashing techniques. Specifically, in this work we focus on cuckoo hashing, a hashing approach introduced by Pagh and Rodler [PR04]. Cuckoo hashing is an efficient dynamic dictionary with highly practical performance guarantees. It provides amortized constant insertion time, worst case constant deletion time and lookup time, and good memory utilization. Additional attractive features of cuckoo hashing are that no dynamic memory allocation is performed, and that the lookup procedure queries only two memory entries which are independent and can be queried in parallel.

Although the insertion time of cuckoo hashing is essentially constant, with a noticeable probability during the insertion of $n$ elements into the hash table, some insertion requires $\Omega(\log n)$ time. Whereas such an amortized performance guarantee is suitable for a wide range of applications, in other applications this is highly undesirable. For these applications, the time per operation must be bounded in the worst case, or at least, the probability that some operation requires a significant amount of time must be negligible. For example, Kirsch and Mitzenmacher [KM07] considered the context of router hardware, where hash tables implementing dynamic dictionaries are used for a variety of operations, including various network measurements and monitoring tasks (see, for example, the work of Broder and Mitzenmacher [BM01] that focuses on the specific task of IP address lookups). In this setting, routers must keep up with line speeds and memory accesses are at a premium.

**Clocked adversaries.** An additional motivation for the construction of dictionaries with worst case guarantees on the time it takes to perform operations was first suggested by Lipton and Naughton [LN93]. One of the basic assumptions in the analysis of probabilistic data structures (first suggested by Carter and Wegman [CW79]) is that the elements that are inserted into the data structure are chosen independently of the randomness used by the data structure. This assumption is violated when the set of elements inserted might be influenced by the time it took the data structure to complete previous operations. Such timing information may reveal sensitive information on the randomness used by the data structure. For example, if the data structure is used for an operating system, then the time a process took to perform an operation affects which process is scheduled and that in turns affects the values of the inserted elements.

This motivates considering "clocked adversaries" – adversaries that can measure the exact time for each operation. Lipton and Naughton actually showed that several dynamic hashing schemes are susceptible to attacks by clocked adversaries, and demonstrated that clocked adversaries can identify elements whose insertion results in poor running time. The concern regarding timing information is even more acute in a cryptographic environment with an active adversary who might use timing information to compromise the system. The adversary might use the timing information to figure out sensitive information on the identity of the elements inserted, or as in the Lipton-Naughton case, to come up with a bad set of elements where even the amortized performance is bad. Note that timing attacks have been shown to be quite powerful and detrimental in cryptography (see, for example, [Koc96, OST06] and the references therein). To combat such attacks, at the very least we want the data structure to devote a fixed amount of time for each operation. There are further concerns like cashing effects, but these are beyond the scope of this paper. Having a fixed upper bound on the time each operation (insert, delete, and lookup) takes, and an exact clock we can, in principle, make the response for each operation be independent of the input and the randomness used.

**Dynamic real-time hashing.** Dietzfelbinger and Meyer auf der Heide [DMadH90] constructed a dynamic dictionary with worst case time per operation and linear space (based on the dynamic dictionary of Dietzfelbinger et al. [DKM$^+$94]). More formally, for any constant $c > 0$ (determined prior to the initialization phase) and for any sequence of operations involving $n$ elements, with probability at least $1 - n^{-c}$ each operation is performed in constant time (that depends on $c$). We are not aware of any other dynamic dictionary with such provable performance guarantees. While the construction of Dietzfelbinger and Meyer auf der Heide is a significant theoretical contribution, it seems that their dictionary may be unsuitable for highly demanding applications. Most notably, it suffers from hidden constant factors in its running time and space utilization, and from an inherently hierarchal structure.

**De-amortized cuckoo hashing.** Motivated by the problem of constructing a *practical* dynamic dictionary with constant worst-case operations, Kirsch and Mitzenmacher [KM07] recently suggested an approach for de-amortizing the insertion time of cuckoo hashing, while essentially preserving the attractive features of the scheme. Specifically, Kirsch and Mitzenmacher suggested an approach for limiting the number of moves per insertion by using a small content-addressable memory (CAM) as a queue for elements being moved. They demonstrated a significant improvement to the worst case performance of cuckoo hashing via experimental results, but left open the problem of constructing a scheme with provable properties.

## 1.1 Our Contributions

In this work we construct the first practical and efficient dynamic dictionary that provably supports constant worst case operations. We follow the approach of Kirsch and Mitzenmacher [KM07] for de-amortizing the insertion time of cuckoo hashing using a queue, while preserving many of the attractive features of the scheme. Specifically, for any polynomial $p(n)$ and constant $\epsilon > 0$ the parameters of our dictionary can be set such that the following properties hold[1]:

1. For any sequence of $p(n)$ insertions, deletions, and lookups, in which at any point in time at most $n$ elements are stored in the data structure, with probability at least $1 - 1/p(n)$ each operation is performed in constant time, where the probability is over the randomness of the initialization phase[2].

2. The memory utilization is essentially 50%. Specifically, the dictionary utilizes $2(1 + \epsilon)n + n^\epsilon$ words.

An additional attractive property is that we never perform rehashing. In general, rehashing is highly undesirable in practice for various reasons, and in particular, it significantly hurts the worst case performance. We avoid rehashing by following the approach of Kirsch, Mitzenmacher and

---

[1] This is the same flavor of worst case guarantee as in the dynamic dictionary of Dietzfelbinger and Meyer auf der Heide [DMadH90].

[2] We note that the non-constant lower bound of Sundar for the membership problem in deterministic dictionaries implies that this type of guarantee is essentially the best possible (see [Sun91], and also the survey of Miltersen [Mil99] who reports on [Sun93]).

Wieder [KMW08] who suggested an augmentation to cuckoo hashing: exploiting a secondary data structure for "stashing" problematic elements that cannot be otherwise stored. We show that in our case, this can be achieved very efficiently by implicitly storing the stash inside the queue.

We provide a formal analysis of the worst-case performance of our dictionary, by generalizing known results in the theory of random graphs. In addition, our analysis involves an application of a recent result due to Braverman [Bra09], to prove that polylog($n$)-wise independent hash functions are sufficient for our dictionary. We note that this is a rather general technique, that may find additional applications in various similar settings. Our extensive experimental results clearly demonstrate that the scheme is highly practical. This seems to be the first dynamic dictionary that simultaneously enjoys all of these properties.

## 1.2 Related Work

Several generalizations of cuckoo hashing circumvent the 50% memory utilization barrier: Fotakis et al. [FPS+05] suggested to use more than two hash functions; Panigraphy [Pan05] and Dietzfelbinger and Weidling [DW07] suggested to store more than one element in each entry. These generalizations led to essentially optimal memory utilization, while preserving the efficiency in terms of update time and lookup time.

Kirsch, Mitzenmacher and Wieder [KMW08] provided an augmentation to cuckoo hashing in order to avoid rehashing. Their idea is to exploit a secondary data structure, referred to as a *stash*, for storing elements that cannot be stored without rehashing. Kirsch et al. proved that for cuckoo hashing with overwhelming probability the number of stashed elements is a very small constant. This augmentation was a crucial ingredient in the work of Naor, Segev, and Wieder [NSW08], who constructed a history independent variant of cuckoo hashing.

Very recently, Dietzfelbinger and Schellbach [DS09] showed that two natural classes of hash functions, the multiplicative class and the class of linear functions over a prime field, lead to large failure probability if applied in cuckoo hashing. This is in contrast to the positive result of Mitzenmacher and Vadhan [MV08], who showed that pairwise independent hash functions are sufficient, provided that the keys are sampled from a block source with sufficient Renyi entropy.

On the experimental side, Ross [Ros06] showed that optimized versions of cuckoo hashing outperform optimized versions of quadratic probing and chained-bucket hashing (the latter is a variant of chained hashing) on the Pentium 4 and Cell processors. Zukowski, Héman and Boncz [ZHB06] compared between cuckoo hashing and chained-bucket hashing on the Pentium 4 and Itanium 2 processors in the context of database workloads, also showing that cuckoo hashing is superior.

## 1.3 Paper Organization

The remainder of this paper is organized as follows. In Section 2 we provide a high-level overview of our construction. In Section 3 we formally describe the data structure. We provide the performance analysis of our dictionary in Section 4. In Section 5 we extend the analysis to hash functions that are polylog($n$)-wise independent. The proof of the main technical lemma underlying our analysis is presented in Section 6. In Section 7 we present experimental results. In Section 8 we discuss concluding remarks and open problems.

## 2  Overview of the Construction

In this section we provide an overview of our construction. We first provide a high-level description of cuckoo hashing, and of the approach of Kirsch and Mitzenmacher [KM07] for de-amortizing it. Then, we present our approach together with the mains ideas underlying its analysis.

**Cuckoo hashing.** Cuckoo hashing uses two tables $T_0$ and $T_1$, each consisting of $r = (1 + \epsilon)n$ entries for some constant $\epsilon > 0$, and two hash functions $h_0, h_1 : \mathcal{U} \to \{0, \ldots, r - 1\}$. An element $x \in \mathcal{U}$ is stored either in entry $h_0(x)$ of table $T_0$ or in entry $h_1(x)$ of table $T_1$, but never in both. The lookup procedure is straightforward: when given an element $x \in \mathcal{U}$, query the two possible memory entries in which $x$ may be stored. The deletion procedure deletes $x$ from the entry in which it is stored. As for insertions, Pagh and Rodler [PR04] proved that the "cuckoo approach", kicking other elements away until every element has its own "nest", leads to a highly efficient insertion procedure. More specifically, in order to insert an element $x \in \mathcal{U}$ we first query entry $T_0[h_0(x)]$. If this entry is not occupied, we store $x$ in that entry. Otherwise, we store $x$ in that entry anyway, thus making the previous occupant "nestless". This element is then inserted to $T_1$ in the same manner, and so forth iteratively. We refer the reader to [PR04] for a more comprehensive description of cuckoo hashing.

**De-amortization using a queue.** Although the amortized insertion time of cuckoo hashing is constant, with a noticeable probability during the insertion of $n$ elements into the hash table, some insertion requires moving $\Omega(\log n)$ elements before identifying an unoccupied entry. We follow the approach of Kirsch and Mitzenmacher [KM07] for de-amortizing cuckoo hashing by using a queue. The main idea underlying the construction of Kirsch and Mitzenmacher is as follows. A new element is always inserted to the queue. Then, an element $x$ is chosen from the queue, according to some queueing policy, and is inserted into the tables. If this is the first insertion attempt for the element $x$ (i.e., $x$ was never stored in one of the tables), then we store it in entry $T_0[h_0(x)]$. If this entry is not occupied, we are done. Otherwise, the previous occupant $y$ of that entry is inserted into the queue, together with an additional information bit specifying that the next insertion attempt for $y$ should begin with table $T_1$. The queueing policy then determines the next element to be chosen from the queue, and so on.

In their experiments, Kirsch and Mitzenmacher loaded the queue with many insert operations, and let the system run. This, however, does not constitute a full description of the insert operation, which should be defined given *one new element*. Specifically, they left unspecified the method of determining the number of operations that are performed upon the insertion of a new element.

**Our approach.** In this work we propose a de-amortization of cuckoo hashing that provably guarantees worst case constant insertion time (with overwhelming probability over the randomness of the initialization phase). Our insertion procedure is parameterized by a constant $L$, and is defined as follows. Given a new element $x \in \mathcal{U}$, we place the pair $(x, 0)$ at the *back* of the queue (the additional bit 0 indicates that the element should be inserted to table $T_0$). Then, we take the pair at the *head* of the queue, denoted $(y, b)$, and place $y$ in entry $T_b[h_b(y)]$. If this entry is not occupied, we again take the pair that is currently stored at the head of the queue, and repeat the same process. If the entry $T_b[h_b(y)]$ is occupied, however, we place its previous occupant $z$ in entry $T_{1-b}[h_{1-b}(z)]$ and so on, as in the above description of cuckoo hashing. After $L$ elements have been moved, we place the current "nestless" element at the *head* of the queue, together with a bit indicating the next table to which it should be inserted, and terminate the insertion procedure.

The deletion and lookup procedures are naturally defined by the property that any element $x$ is stored in one of $T_0[h_0(x)]$ and $T_1[h_1(x)]$, or in the queue. However, unlike the classical cuckoo hashing, here it is not clear that these procedures run in constant time. It may be the case that the insertion procedure causes the queue to contain many elements, and then the deletion and lookup procedures of the queue will require a significant amount of time.

The main property underlying our construction is that the constant $L$ (i.e., the number of iterations of the insertion procedure) can be chosen such that with overwhelming probability the queue does not contain more than a logarithmic number of elements at any point in time. In this case we show that simple and efficient instantiations of the queue can indeed support insertions, deletions and lookups in worst case constant time. This is proved by considering the distribution of the *cuckoo graph*, formally defined as follows:

**Definition 2.1.** Given a set $S \subseteq \mathcal{U}$ and two hash functions $h_0, h_1 : \mathcal{U} \to \{0, \ldots, r-1\}$, the *cuckoo graph* is the bipartite graph $G = (L, R, E)$, where $L = R = \{0, \ldots, r-1\}$ and $E = \{(h_0(x), h_1(x)) : x \in S\}$.

The main idea of our analysis is to consider $\log n$ insertions each time, and to examine the total number of moves in the cuckoo graph that these $\log n$ insertions require. Our main technical contribution in this setting is proving that the sum of sizes of any $\log n$ connected components in the cuckoo graph is upper bounded by $O(\log n)$ with overwhelming probability. This is a generalization of a well-known bound in graph theory on the size of a single connected component.

**Avoiding rehashing.** It is rather easy to see that a set $S$ can be successfully stored in the cuckoo graph using hash functions $h_0$ and $h_1$ if and only if no connected component in the graph has more edges then nodes. In other words, every component contains at most one cycle (unicyclic). It is known, however, that even if $h_0$ and $h_1$ are completely random functions, then with probability $\Theta(1/n)$ there will be a connected component with more than one cycle. In this case the given set cannot be stored using $h_0$ and $h_1$. The standard solution for this scenario is to choose new functions and rehash the entire data. This significantly hurts the worst case performance of the data structure (and is highly undesirable in practice for various other reasons).

To overcome this difficulty, we follow the approach of Kirsch et al. [KMW08] who suggested an augmentation to cuckoo hashing in order to avoid rehashing: exploiting a secondary data structure, referred to as a *stash*, for storing elements that create cycles, starting from the second cycle of each component. That is, whenever an element is inserted into a unicyclic component and creates an additional cycle in this component, the element is stashed. Kirsch et al. showed that this approach performs remarkably well by proving that for any fixed set $S$ of size $n$, the probability that at least $k$ elements are stashed is $O(n^{-k})$ (see Lemma 4.5 in Appendix 4). In our setting, however, where the data structure has to support delete operations in constant time, it is not straightforward to use a stash explicitly. Specifically, for the stash to remain of constant size, after every delete operation it may be required to move some element back from the stash to one of the two tables. Otherwise, the analysis of Kirsch et al. on the size of the stash no longer holds when considering long sequences of operations on the data structure.

We overcome this difficulty by storing the stashed elements in the queue. That is, whenever we identify an element that closes a second cycle in the cuckoo graph, this element is placed at the back of the queue. Very informally, this guarantees that any stashed element is given a chance to be inserted back to the tables after essentially $\log n$ invocation of the insertion procedure. This implies that the number of stashed elements in the queue roughly corresponds to the number of elements that close a second cycle in the cuckoo graph at any point in time (up to intervals of $\log n$ insertions). We can then use the result of Kirsch et al. [KMW08] to argue that there is a very small number of such elements in the queue at any point.

For detecting cycles in the cuckoo graph we implement a simple *cycle detection mechanism* (CDM), as suggested by Kirsch et al. [KMW08]. When inserting an element we insert to the CDM all the elements that are encountered in its connected component during the insertion process. Once we identify that a component has more than one cycle we stash the current nestless element (i.e., place it in the back of the queue), and reset the CDM to its initial configuration. We note that in the classical cuckoo hashing cycles are detected by allowing the insertion procedure to run for $O(\log n)$ steps, and then announcing failure (which is followed by rehashing). In our case, however, it is crucial that a cycle is detected in time that is linear in the size of its connected component in the cuckoo graph.

**Using polylog(n)-wise independent hash functions.** When analyzing the performance of our scheme, we first assume the availability of truly random hash functions. Then, we apply a recent result of Braverman [Bra09] and show that the same performance guarantees hold when instantiating our scheme with hash functions that are only polylog($n$)-wise independent (see [DW03, OP03, Sie89] for efficient constructions of such functions with succinct representations and constant evaluation time). Informally, Braverman proved that for any Boolean circuit $C$ of depth $d$, size $m$, and unbounded fan-in, and for any $k$-wise distribution $X$ with $k = (\log m)^{O(d^2)}$, it holds that $\mathbb{E}[C(U_n)] \approx \mathbb{E}[C(X)]$. That is, $X$ "fools" the circuit $C$ into behaving as if $X$ is the uniform distribution $U_n$ over $\{0,1\}^n$.

Specifically, in our analysis we define a "bad" event with respect to the hash values of $h_0$ and $h_1$, and prove that: (1) this event occurs with probability at most $n^{-c}$ (for an arbitrarily large constant $c$) assuming truly random hash functions, and (2) as long as this event does not occur each operation is performed in constant time. We show that this event can be recognized by a Boolean circuit of constant depth, size $m = n^{O(\log n)}$, and unbounded fan-in. In turn, Braverman's result implies that it suffices to use $k$-wise independent hash functions for $k = \text{polylog}(n)$.

We note that applying Braverman's result in such setting is quite a general technique and may be found useful in other similar scenarios. In particular, our argument implies that the same holds for the analysis of Kirsch et al. [KMW08], who proved the above-mentioned bound on the number of stashed elements assuming that the underlying hash functions are truly random.

## 3 The Data Structure

As discussed in Section 2, our data structure uses two tables $T_0$ and $T_1$, and two auxiliary data structures: a queue, and a cycle-detection mechanism. Each table consists of $r = (1+\epsilon)n$ entries for some small constant $\epsilon > 0$. Elements are inserted into the tables using two hash functions $h_0, h_1 : \mathcal{U} \to \{0, \ldots, r-1\}$, which are independently chosen at the initialization phase. We assume that the auxiliary data structures satisfy the following properties (we emphasize that these data structures will contain a very small number of elements with overwhelming probability, and in Section 3.1 we propose simple instantiations):

1. The queue is constructed to store at most $O(\log n)$ elements at any point in time. It should support the operations `Lookup`, `Delete`, `PushBack`, `PushFront`, and `PopFront` in worst-case constant time (with overwhelming probability over the randomness of its initialization phase).

2. The cycle-detection mechanism is constructed to store at most $O(\log n)$ elements at any point in time. It should support the operations `Lookup`, `Insert` and `Reset` in worst-case constant time (with overwhelming probability over the randomness of its initialization phase).

An element $x \in \mathcal{U}$ can be stored in exactly one out of three possible places: entry $h_0(x)$ of table $T_0$, entry $h_1(x)$ of table $T_1$, or the queue. The lookup procedure is straightforward: when given an element $x \in \mathcal{U}$, query the two tables and if needed, perform lookups in the queue. The deletion procedure is also straightforward by first searching for the element, and then deleting it. The insertion procedure was essentially already described in Section 2. A formal description of these procedures is provided in Figure 2 and a schematic diagram of the whole data structure is presented in Figure 1.

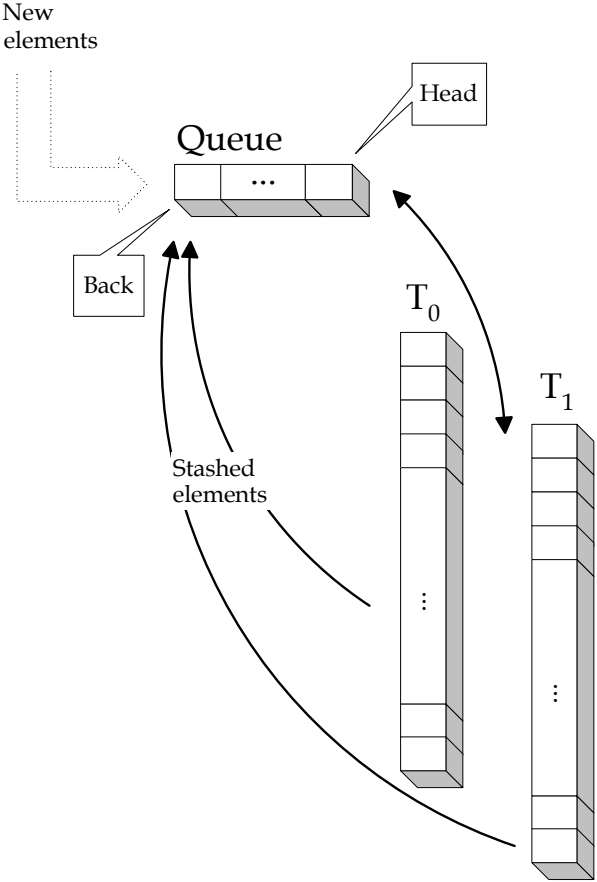

**Figure 1:** A schematic diagram of our dictionary.

```
Initialize():
 1: for i = 0 to r − 1 do
 2:     T_0[i] ← ⊥
 3:     T_1[i] ← ⊥
 4: InitializeQueue()
 5: InitializeCDM()

Lookup(x):
 1: if T_0[h_0(x)] = x or T_1[h_1(x)] = x then
 2:     return true
 3: if LookupQueue(x) then
 4:     return true
 5: return false

Delete(x):
 1: if T_0[h_0(x)] = x then
 2:     T_0[h_0(x)] ← ⊥
 3:     return
 4: if T_1[h_1(x)] = x then
 5:     T_1[h_1(x)] ← ⊥
 6:     return
 7: DeleteFromQueue(x)

Insert(x):
 1: InsertIntoBackOfQueue(x, 0)
 2: y ← ⊥  // y denotes the current element we work with
 3: for i = 1 to L do
 4:     if y = ⊥ then  // Fetching element y from the head of the queue
 5:         if IsQueueEmpty() then
 6:             return
 7:         else
 8:             (y, b) ← PopFromQueue()
 9:     if T_b[h_b(y)] = ⊥ then  // Successful insert
10:         T_b[h_b(y)] ← y
11:         ResetCDM()
12:         y ← ⊥
13:     else
14:         if LookupInCDM(y, b) then  // Found the second cycle
15:             InsertIntoBackOfQueue(y, b)
16:             ResetCDM()
17:             y ← ⊥
18:         else  // Evict existing element
19:             z ← T_b[h_b(y)]
20:             T_b[h_b(y)] ← y
21:             InsertIntoCDM(y, b)
22:             y ← z
23:             b ← 1 − b
24: if y ≠ ⊥ then
25:     InsertIntoHeadOfQueue(y, b)
```

**Figure 2:** The Initialize, LookUp, Delete and Insert procedures.

In Section 4 we analyze the performance of the data structure, and prove the following theorem:

**Theorem 3.1.** *For any polynomial $p(n)$ and constant $\epsilon > 0$, the parameters of the dictionary can be set such that the following properties hold:*

1. *For any sequence of at most $p(n)$ insertions, deletions, and lookups, in which at any point in time at most $n$ elements are stored in the dictionary, with probability at least $1 - 1/p(n)$ each operation is performed in constant time, where the probability is over the randomness of the initialization phase.*

2. *The dictionary utilizes $2(1 + \epsilon)n + n^\epsilon$ words.*

### 3.1 The Auxiliary Data Structures

We propose simple instantiations for the auxiliary data structures. Any other instantiations that satisfy the above-mentioned properties are also possible.

**The queue.** In Section 4 we will argue that with overwhelming probability the queue contains at most $O(\log n)$ elements at any point in time. Therefore, we design the queue to store at most $O(\log n)$ elements, and allow the whole data structure to fail if the queue overflows. Although a classical queue can support the operations PushBack, PushHead, and PopFront in constant time, we also need to support the operations Lookup and Delete in constant time. One possible instantiation is to use a constant number $k$ arrays $A_1, \ldots, A_k$ each of size $n^\delta$, for some $\delta < 1$. Each entry of these arrays consists of a data element, a pointer to the previous element in the queue, and a pointer to the next element in the queue. In addition we maintain two global pointers: the first points to the head of the queue, and the second points to the end of the queue. The elements are stored using a function $h$ chosen from a collection of pairwise independent hash functions. Specifically, each element $x$ is stored in the first available entry amongst $\{A_1[h(1,x)], \ldots, A_k[h(k,x)]\}$. For any element $x$, the probability that all of its $k$ possible entries are occupied when the queue contains at most $m = O(\log n)$ elements is upper bounded by $(m/n^\delta)^k$, which can be made as small as $n^{-c}$ for any constant $c$ by choosing an appropriate constant $k$.

**The cycle-detection mechanism.** As in the case of the queue, in Section 4 we will argue that with overwhelming probability the cycle-detection mechanism contains at most $O(\log n)$ element at any point in time. Therefore, we design the cycle-detection mechanism to store at most $O(\log n)$ elements, and allow the whole data structure to fail if the cycle-detection mechanism overflows. One possible instantiation is to use the above-mentioned instantiation of the queue together with any standard augmentation that enables constant time resets (see, for example, [BT93]).

## 4 Performance Analysis

In this section we prove Theorem 3.1. In terms of memory utilization, each of the two tables $T_0$ and $T_1$ has $(1 + \epsilon)n$ entries, and the auxiliary data structures (as suggested in Section 3.1) require sublinear space. Therefore, the memory utilization is essentially 50%, as in the standard cuckoo hashing. In terms of running time, we say that the auxiliary data structures *overflow* if either the queue or the cycle-detection mechanism contain more than $O(\log n)$ elements. We show that as long as the auxiliary data structures do not fail or overflow, all operations are performed in constant time. As suggested in Section 3.1, for any constant $c$ the auxiliary data structures can be constructed such that they fail with probability less than $n^{-c}$, and therefore we only need to bound the probability of overflow. We deal with each of the auxiliary data structures separately. For the remainder of the analysis we introduce the following definition and notation:

**Definition 4.1.** A sequence $\pi$ of insert, delete and lookup operations is *n-bounded* if at any point in time during the execution of $\pi$ the data structure contains at most $n$ elements.

**Notation 4.2.** For an element $x \in \mathcal{U}$ we denote by $C_{S,h_0,h_1}(x)$ the connected component that contains the edge $(h_0(x), h_1(x))$ in the cuckoo graph of the set $S \subseteq \mathcal{U}$ with functions $h_0$ and $h_1$.

We prove the following theorem:

**Theorem 4.3.** *For any polynomial $p(n)$ and any constant $\epsilon > 0$, there exists a constant $L$ such that when instantiating the data structure with parameters $\epsilon$ and $L$ the following holds: For any n-bounded sequence of operations $\pi$ of length at most $p(n)$, with probability $1 - 1/p(n)$ over the coin tosses of the initialization phase the auxiliary data structures do not overflow during the execution of $\pi$.*

We define two "good" events, and show that as long as these events occur, then the queue and the cycle-detection mechanism do not overflow. Let $\pi$ be an $n$-bounded sequence of $p(n)$ operations. Denote by $(x_1, \ldots, x_N)$ the elements inserted by $\pi$ in reverse order. Note that between any two insertions $\pi$ may perform several deletions, and therefore an element may appear more than once. For any integer $1 \leq j \leq N/\log n$, denote by $S_j$ the set of elements that are stored in the data structure just before the insertion of $x_{(j-1)\log n+1}$, together with the elements $\{x_{(j-1)\log n+1}, \ldots, x_{j\log n}\}$. That is, the set $S_j$ contains the result of executing $\pi$ up to $x_{j\log n}$ while ignoring any deletions that occur between $x_{(j-1)\log n}$ and $x_{j\log n}$. Note that since $\pi$ is an $n$-bounded sequence, we have that $|S_j| \leq n + \log n$ for all $j$'s. In Section 6 we prove the following lemma, which is the main technical tool in the proof of Theorem 4.3:

**Lemma 4.4.** *For any constants $\epsilon, c_1 > 0$ and any integer $T \leq \log n$ there exists a constant $c_2$, such that for any set $S \subseteq \mathcal{U}$ of size $n$ and for any $x_1, \ldots, x_T \in S$ it holds that*

$$\Pr\left[\sum_{i=1}^{T} |C_{S,h_0,h_1}(x_i)| \geq c_2 T\right] \leq \exp(-c_1 T) \;,$$

*where the probability is taken over the random choice of the functions $h_0, h_1 : \mathcal{U} \to \{0, \ldots, r-1\}$, for $r = (1+\epsilon)n$.*

We now define the good events. Denote by $\mathcal{E}_1$ the event in which for every $1 \leq j \leq N/\log n$ it holds that

$$\sum_{i=1}^{\log n} |C_{S_j,h_0,h_1}(x_{(j-1)\log n+i})| \leq c_2 \log n \;.$$

An appropriate choice of the constant $c_1$ in Lemma 4.4 and a union bound imply that the event $\mathcal{E}_1$ occurs with probability at least $1 - n^{-c}$. A minor technical detail is that Lemma 4.4 is stated for sets $S$ of size at most $n$ (for simplicity), whereas the $S_j$'s are of size at most $n + \log n$. This, however, is easily fixed by replacing $\epsilon$ with $\epsilon' = 2\epsilon$ in the statement of the lemma.

In addition, denote by $\mathsf{stash}(S_j, h_0, h_1)$ the number of stashed elements (as discussed in Section 2) in the cuckoo graph of $S_j$ with hash functions $h_0$ and $h_1$. Denote by $\mathcal{E}_2$ the event in which for every $1 \leq j \leq N/\log n$ it holds that $\mathsf{stash}(S_j, h_0, h_1) \leq k$. The following lemma of Kirsch et al. [KMW08] implies that the constant $k$ can be chosen such that the probability of the event $\mathcal{E}_2$ is at least $1 - n^{-c}$ (we note that the above comment on $n$ vs. $n + \log n$ holds here as well).

**Lemma 4.5** ([KMW08]). *For any set $S \subseteq \mathcal{U}$ of size $n$, the probability that the stash contains at least $k$ elements is $O(n^{-k})$, where the probability is taken over the random choice of the functions $h_0, h_1 : \mathcal{U} \to \{0, \ldots, r-1\}$, for $r = (1+\epsilon)n$.*

The following claims prove Theorem 4.3:

**Claim 4.6.** *Let $\pi$ be an $n$-bounded sequence of $p(n)$ operations. Assuming that the events $\mathcal{E}_1$ and $\mathcal{E}_2$ occur, then during the execution of $\pi$ the queue does not contain more than $2\log n + k$ elements at any point in time.*

**Claim 4.7.** *Let $\pi$ be an $n$-bounded sequence of $p(n)$ operations. Assuming that the events $\mathcal{E}_1$ and $\mathcal{E}_2$ occur, then during the execution of $\pi$ the cycle-detection mechanism does not contain more than $(c_2 + 1) \log n$ elements at any point in time.*

**Proof of Claim 4.6.** We prove by induction on $j$, that at the time $x_{j \log n + 1}$ is inserted into the queue, there are no more than $\log n + k$ elements in the queue. This clearly implies that at any point in time there are at most $2 \log n + k$ elements in the queue.

For $j = 1$ we observe that there are at most $\log n$ elements in the data structure at that point in time. In particular, there are at most $\log n$ elements in the queue.

Assume that the statement holds for some $j$, and we prove that it holds also for $j + 1$. The inductive hypothesis states that at the time $x_{j \log n + 1}$ is inserted, the queue contains at most $\log n + k$ elements. In the worst case, these elements are $\{x_{(j-1) \log n + 1}, \ldots, x_{j \log n}\}$ together with some additional $k$ elements. It is rather straightforward that the number of moves in the cuckoo graph that are required for inserting an element is at most the size of its connected component. Therefore, the event $\mathcal{E}_1$ implies that the elements $\{x_{(j-1) \log n + 1}, \ldots, x_{j \log n}\}$ can be inserted in $c_2 \log n$ moves, and that each of the additional $k$ elements can be inserted in at most $c_2 \log n$ moves. Therefore, these $\log n + k$ elements can be inserted in $c_2 \log n + k c_2 \log n$ moves. By choosing the constant $L$ such that $L \log n \geq c_2 \log n + k c_2 \log n$ it is guaranteed that by the time the element $x_{(j+1) \log n + 1}$ is inserted, these $\log n + k$ elements will be inserted into the tables, and at most $k$ of them with be stored in the queue due to second cycles in the cuckoo graph (due to event $\mathcal{E}_2$). Thus, by the time the element $x_{(j+1) \log n + 1}$ is inserted, the queue contains (in the worst case) the elements $\{x_{j \log n + 1}, \ldots, x_{(j+1) \log n}\}$, and some additional $k$ elements. ∎

**Proof of Claim 4.7.** At any point in time the cycle-detection mechanism contains elements from exactly one connected component in the cuckoo graph. Therefore, at any point in time the number of elements stored in the cycle-detection mechanism is at most the number of element in the maximal connected component. The event $\mathcal{E}_1$ guarantees that there is no set $S_j$ with a connected component containing more than $c_2 \log n$ elements. Between the $S_j$'s, at most $\log n$ elements are inserted, and this guarantees that at any point in time the cycle-detection mechanism does not contain more than $(c_2 + 1) \log n$ elements. ∎

## 5 Using polylog($n$)-wise Independent Hash Functions

The proof provided in Section 4 for our main theorem relies on Lemmata 4.4 and 4.5 on the structure of the cuckoo graph. These lemmata are the only part of the proof in which the amount of independence of the hash functions $h_0$ and $h_1$ is taken into consideration. Briefly, Lemma 4.4 states that the probability that the sum of sizes of $T$ connected components in the cuckoo graph exceeds $cT$ is exponentially small (for $T \leq \log n$), and in Section 6 we prove this lemma for truly random hash functions. Lemma 4.5 states that for any set of $n$ elements, the probability that the stash contains at least $k$ elements is $O(n^{-k})$, and this Lemma was proved by Kirsch et al. [KMW08] for truly random hash functions.

In this section we show that these two lemmata hold even if $h_0$ and $h_1$ are sampled from a family of polylog($n$)-wise independent hash functions. We apply a recent result of Braverman [Bra09] (which is a significant extension of prior work of Bazzi [Baz07] and Razborov [Raz08]), and note that this approach is quite a general technique and may be found useful in other similar scenarios. Braverman proved that for any Boolean circuit $C$ of depth $d$, size $m$, and unbounded fan-in, and for any $r$-wise distribution $X$ with $r = (\log m)^{O(d^2)}$, it holds that $\mathbb{E}[C(U_n)] \approx \mathbb{E}[C(X)]$. That is, $X$ "fools" the circuit $C$ into behaving as if $X$ is the uniform distribution $U_n$ over $\{0,1\}^n$. More formally, Braverman proved the following theorem:

**Theorem 5.1** ([Bra09]). *Let $s \geq \log m$ be any parameter. Let $F$ be a boolean function computed by a circuit of depth $d$ and size $m$. Let $\mu$ be an $r$-independent distribution where*

$$r \geq 3 \cdot 60^{d+3} \cdot (\log m)^{(d+1)(d+3)} \cdot s^{d(d+3)} \ ,$$

*then*

$$|\mathbb{E}_\mu[F] - \mathbb{E}[F]| < \varepsilon(s,d) \ ,$$

*where $\varepsilon(s,d) = 0.82^s \cdot 15m$.*

In our analysis in Section 4 we defined two "bad" events with respect to the hash values of $h_0$ and $h_1$. The first event corresponds to Lemma 4.4 and the second event corresponds to Lemma 4.5:

**Event 1:** There exists a set $S$ of $T \leq \log n$ vertices in the cuckoo graph, such that the sum of sizes of the connected components of the vertices in $S$ is larger than $cT$, for some constant $c$ (this is the complement of the event $\mathcal{E}_1$ defined in Section 4).

**Event 2:** There exists a set $S$ of at most $n$ vertices in the cuckoo graph, such that the number of stashed elements from the set $S$ exceeds some constant $k$ (this is the complement of the event $\mathcal{E}_2$ defined in Section 4).

In what follows we show that these two events can be recognized by constant-depth and quasi-polynomial size Boolean circuits, which will enable us to apply Theorem 5.1 to get the desired result. The input wires of our circuits contain the values $h_0(x_1), h_1(x_1), \ldots, h_0(x_n), h_1(x_n)$ (where the $x_i$'s represent the elements inserted into the data structure).

**Identifying event 1.** This event occurs if and only if the graph contains at least one forest from a specific set of forests of the bipartite graph on $[r] \times [r]$, where $r = (1+\epsilon)n$. We denote this set of forests by $\mathcal{F}_n$, and observe that $\mathcal{F}_n$ is a subset of all forests with at most $cT + 1 = O(\log n)$ vertices, which implies that $|\mathcal{F}_n| = n^{O(\log n)}$. Therefore, the event can be identified by a constant-depth circuit of size $n^{O(\log n)}$ that simply enumerates all forests $F \in \mathcal{F}_n$, and for every such forest $F$ the circuit checks whether it exists in the graph:

$$\bigvee_{F \in \mathcal{F}_n} \bigwedge_{(u,v) \in F} \bigvee_{i=1}^{n} \left[ \Big(h_0(x_i) = u \wedge h_1(x_i) = v\Big) \vee \Big(h_0(x_i) = v \wedge h_1(x_i) = u\Big) \right] \quad (5.1)$$

**Identifying event 2.** For identifying this event we go over all subsets $S' \subseteq \{x_1, \ldots, x_n\}$ of size $k$, and for every such subset we check whether all of its elements are stashed. Note, however, that the set of stashed elements is not uniquely defined: given a connected component with two cycles, any element on one of the cycles can be stashed. Therefore, we find it natural to define a "canonical" set of stashed elements, as suggested by Naor et al. [NSW08]: given a connected component with more than one cycle we iteratively stash the largest edge that lies in a cycle (according to some ordering), until the component contains only one cycle[3]. Specifically, given an element $x \in S'$ we enumerate over all connected components in which the edge $(h_0(x), h_1(x))$ is stashed according to the canonical rule above, and check whether the component exists in the cuckoo graph (as in Equation (5.1)). Note that $k$ is constant and that we can restrict ourselves to connected components with $O(\log n)$ vertices (since we can assume that event 1 above does not occur). Therefore, the resulting circuit is of constant depth and size $n^{O(\log n)}$.

## 6 Proof of Lemma 4.4 – Size of Connected Components

In this section we prove Lemma 4.4 that states a property on the structure of the cuckoo graph. Specifically, we are interested in bounding the sum of sizes of several connected components in the graph.

Recall (Definition 2.1) that the cuckoo graph is a bipartite graph $G = (L, R, E)$ with $L = R = [n]$ and $(1 - \epsilon)n$ edges, where each edge is chosen independently and uniformly at random from the set $L \times R$ (note that the number of distinct edges may be less than $(1 - \epsilon)n$). The distribution of the cuckoo graph is very close (in some sense that will be formalized later on) to the well-studied distribution $\mathbb{G}(n, n, M)$ on bipartite graphs $G = (L, R, E)$ where $L = R = [n]$ and $E$ is a set of exactly $M = (1-\epsilon)n$ edges that is chosen uniformly at random from all subsets of size $M$ of $L \times R$. Thus, the proof essentially reduces to consider the random graph model $\mathbb{G}(n, n, M)$.

Our proof is a generalization of a well-known proof for the size of a *single* connected component (see, for example, [JŁR00, Theorem 5.4]). We first prove a similar lemma for the distribution $\mathbb{G}(n, n, p)$ on bipartite graphs $G = ([n], [n], E)$ where each edge is independently chosen with probability $p$ (Section 6.1). Then we apply a standard argument to show that the same holds for the distribution $\mathbb{G}(n, n, M)$ (Section 6.2). Finally, we prove that the lemma holds for the distribution of the cuckoo graph as well.

### 6.1 The $\mathbb{G}(n, n, p)$ Case

In what follows, given a graph $G$ and a vertex $v$ we denote by $C_G(v)$ the connected component of $v$ in $G$. We prove the following lemma:

**Lemma 6.1.** *Let $np = c$ for some constant $0 < c < 1$. For any integer $T \leq \log n$ and any constant $c_1 > 0$ there exists a constant $c_2$, such that for any vertices $v_1, \ldots, v_T \in L \cup R$*

$$\Pr\left[\sum_{i=1}^{T} |C_G(v_i)| \geq c_2 T\right] \leq \exp(-c_1 T) ,$$

*where the graph $G = (L, R, E)$ is sampled from $\mathbb{G}(n, n, p)$.*

We begin by proving a slightly weaker claim that bounds the size of the *union* of several connected components:

**Lemma 6.2.** *Let $np = c$ for some constant $0 < c < 1$. For any integer $T \leq \log n$ and any constant $c_1 > 0$ there exists a constant $c_2$, such that for any vertices $v_1, \ldots, v_T \in L \cup R$*

$$\Pr\left[\left|\bigcup_{i=1}^{T} C_G(v_i)\right| \geq c_2 T\right] \leq \exp(-c_1 T) ,$$

*where the graph $G = (L, R, E)$ is sampled from $\mathbb{G}(n, n, p)$.*

---
[3]We emphasize that this is only for simplifying the analysis, and not to be used by the actual data structure.

**Proof.** We will look at the random graph process that led to our graph $G$. Specifically, we will focus on the set $S = \{v_1, \ldots, v_T\}$ and analyze the process of growth of the components $C_G(v_1), C_G(v_2), \ldots, C_G(v_T)$ as $G$ evolves.

Informally, we divide the evolution process of $G$ into layers (where in every layer there is a number of steps). The first layer consists of the vertices in $S$. The second layer consists of all the vertices in $G$ that are the neighbors of the vertices in $S$. In general, the $i^{th}$ layer consists of vertices that are neighbors of all the vertices in the previous layer $i-1$ (excluding the vertices in layer $i-2$, which are also the neighbors of layer $i-1$, but were already accounted for). This is, in fact, the BFS tree of the graph $H \triangleq \bigcup_{i=1}^{T} C_G(v_i)$. See Figure 3.

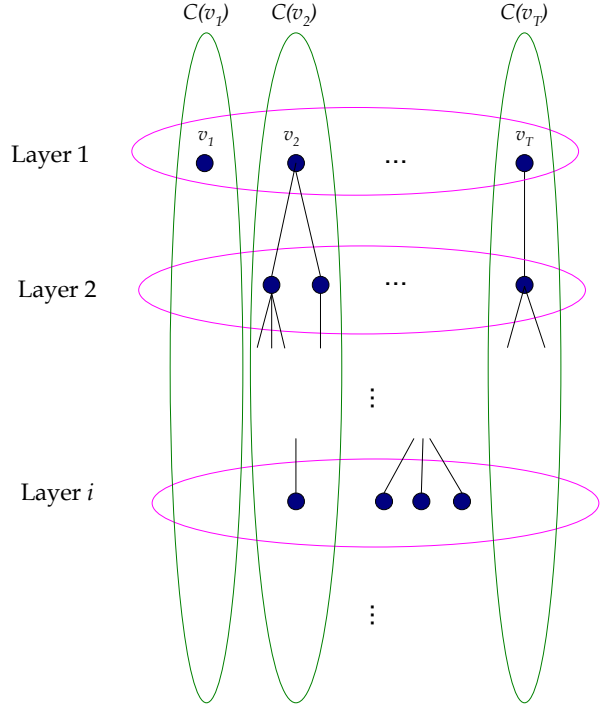

**Figure 3:** The layered view of our random graph evolution.

So instead of developing each of the connected components $C_G(v_i)$ separately, edge by edge, we do it "in parallel" – layer by layer. We start with the set of $T$ isolated vertices, which was denoted by $S$, and develop the neighbors of every $v_i$ in turn: first we connect to $v_1$ its neighbors in $C_G(v_1)$ (if they exist), then we connect to $v_2$ its neighbors in $C_G(v_2)$ (if they exist), and so on, concluding with the neighbors of $v_T$. At this point we are through with layer 2. Next, we go over the vertices in layer 2, connecting each of them with their respective neighbors in layer 3. We finish the whole process when we discover all of the vertices in $H$.

Whenever we add an edge to some vertex in the above process, there are two possibilities:

1. The added edge had contributed an additional vertex to exactly one of the components $C_G(v_1), C_G(v_2), \ldots, C_G(v_T)$. In this case we will add this vertex and increase the size of the appropriate $C_G(v_i)$ by one.

2. The added edge merged any two of the components $C_G(v_i)$ and $C_G(v_j)$. In this case we will merge these components into a single component $C_G(v_i)$ (without loss of generality) and forget about $C_G(v_j)$. Note that this means that we count each vertex in the set $H$ exactly once.

So formally, let us denote for every step $i$ in the process

$$X_i \triangleq \{\text{The number of new vertices we add in step } i\}$$

$X_i$ is dominated by a random variable $Y_i$, where all $Y_i$ have the same binomial distribution $Bi(n,p)$. $Y_i$ are independent, since we never develop two components with a common edge in the same step $\ell$, but rather merge them (if the common edge had appeared in some step $j < \ell$, the two components would have been merged in step $j$, and if the common edge had appeared in step $\ell$, then until that moment (inclusive) $Y_\ell$ and $Y_j$ still behave as two independent binomial variables).

The stochastic process described above is known as *Galton-Watson* process. The probability that a given vertex $v$ belongs to a component of size at least $k = k(n)$ is bounded from above by the probability that the sum of $k$ random variables $Y_i$ is at least $k - T$. Formally,

$$\Pr\left[|H| \geq k\right] \leq \Pr\left[\sum_{i=1}^{k} Y_i \geq k - T\right]$$

In our case $k = c_2 T$, and since $\sum_{i=1}^{c_2 T} Y_i \sim Bi(c_2 T n, p)$, the expectation of this sum is $c_2 T n p = c_2 T c$, due to the assumption $np = c$. Consequently, we need to bound the following deviation from the mean:

$$\Pr\left[\sum_{i=1}^{c_2 T} Y_i \geq c_2 T - T\right] = \Pr\left[\sum_{i=1}^{c_2 T} Y_i \geq c_2 T c + (1-c)c_2 T - T\right]$$

This deviation is positive if we choose $c_2 > \frac{1}{1-c}$. Using Chernoff bound we obtain:

$$\Pr\left[\sum_{i=1}^{c_2 T} Y_i \geq c_2 T c + (1-c)c_2 T - T\right] \leq \exp\left(-\frac{((1-c)c_2 T - T)^2}{2\left(c_2 T c + \frac{(1-c)c_2 T - T}{3}\right)}\right)$$

An easy calculation shows that

$$\exp\left(-\frac{((1-c)c_2 T - T)^2}{2\left(c_2 T c + \frac{(1-c)c_2 T - T}{3}\right)}\right) \leq \exp(-c_1 T) ,$$

when choosing $c_2 \geq \max\left\{\frac{1}{1-c}, \frac{c_1 + 3/2}{3c + 3/2}\right\}$. ∎

**Proof of Lemma 6.1.** In general, a bound on $\left|\bigcup_{i=1}^{T} C_G(v_i)\right|$ does not imply a similar bound on $\sum_{i=1}^{T} |C_G(v_i)|$. In our case, however, for $T \leq \log n$ we can argue that with high probability the two are related up to a constant multiplicative factor.

For a constant $c_3$, we denote by $\mathcal{S}_{ame(c_3)}$ the event in which some $c_3$ vertices from the set $\{v_1, \ldots, v_T\}$ are in the same connected component. Then,

$$\Pr\left[\mathcal{S}_{ame(c_3)} \middle| \left|\bigcup_{i=1}^{T} C_G(v_i)\right| \leq c_2 T\right] \leq T \cdot \binom{T}{c_3} \cdot \left(\frac{c_2 T}{n}\right)^{c_3}$$

$$\leq \frac{(c_2 e)^{c_3} \cdot T^{2c_3 + 1}}{c_3^{c_3} \cdot n^{c_3}}$$

(6.1)

In the right-hand side of the first inequality, the first term comes from the union bound on the $T$ components, the second term counts the number of possible arrangements of the $c_3$ vertices inside the component, and the last term is the probability that all the $c_3$ vertices fall into this specific component. In addition,

$$\Pr\left[\sum_{i=1}^{T}|C_G(v_i)| > c_2 c_3 T\right] \leq \Pr\left[\left|\bigcup_{i=1}^{T} C_G(v_i)\right| > c_2 T \bigvee \mathcal{S}ame(c_3)\right]$$

$$\leq \Pr\left[\left|\bigcup_{i=1}^{T} C_G(v_i)\right| > c_2 T\right] + \Pr\left[\mathcal{S}ame(c_3) \middle| \left|\bigcup_{i=1}^{T} C_G(v_i)\right| \leq c_2 T\right] \quad (6.2)$$

Combining the result of Lemma 6.2 with (6.1) yields:

$$(6.2) \leq \exp(-c_1 T) + \frac{(c_2 e)^{c_3} \cdot T^{2c_3+1}}{c_3^{c_3} \cdot n^{c_3}} \quad (6.3)$$

Therefore, for $T \leq \log n$ there exist constants $c_4$ and $c_5$ such that

$$(6.3) \leq \exp(-c_1 T) + \exp(-c_4 T) \leq \exp(-c_5 T)$$

∎

## 6.2 The $\mathbb{G}(n, n, M)$ Case

The following claim is a straightforward generalization of the well-known relationship between $\mathbb{G}(n, p)$ and $\mathbb{G}(n, M)$ (see, for example, [Bol85, Theorem II.2]):

**Lemma 6.3.** *Let $\mathcal{Q}$ be any graph property and suppose $0 < p = M/n^2 < 1$. Then,*

$$\Pr_{\mathbb{G}(n,n,M)}[\mathcal{Q}] \leq e^{1/(6M)}\sqrt{2\pi p(1-p)n^2} \Pr_{\mathbb{G}(n,n,p)}[\mathcal{Q}]$$

**Proof.** For any graph property $\mathcal{Q}$ the following holds:

$$\Pr_{\mathbb{G}(n,n,p)}[\mathcal{Q}] = \sum_{m=0}^{n^2} \Pr_{\mathbb{G}(n,n,m)}[\mathcal{Q}] \cdot \binom{n^2}{m} p^m (1-p)^{n^2-m}$$

and by fixing $m = M$ we get:

$$\Pr_{\mathbb{G}(n,n,p)}[\mathcal{Q}] \geq \Pr_{\mathbb{G}(n,n,M)}[\mathcal{Q}] \cdot \binom{n^2}{M} p^M (1-p)^{n^2-M} \quad (6.4)$$

By using the following inequality (for instance, cf. [Bol85, inequality (5) in I.1])

$$\binom{n^2}{M} \geq \frac{1}{e^{1/(6M)}\sqrt{2\pi}} \left(\frac{n^2}{M}\right)^M \left(\frac{n^2}{n^2 - M}\right)^{n^2-M} \sqrt{\frac{n^2}{M(n^2-M)}}$$

and the fact that $p = M/n^2$ we get:

$$(6.4) \geq \frac{\Pr_{\mathbb{G}(n,n,M)}[\mathcal{Q}]}{e^{1/(6M)}\sqrt{2\pi p(1-p)n^2}}$$

∎

Lemma 6.1 and Lemma 6.3 yield the following corollary:

**Corollary 6.4.** *Fix $n$ and $M$ such that $M < n$. For any integer $T \leq \log n$ and any constant $c_1 > 0$ there exists a constant $c_2$, such that for any vertices $v_1, \ldots, v_T \in L \cup R$*

$$\Pr\left[\sum_{i=1}^{T} |C_G(v_i)| \geq c_2 T\right] \leq \exp(-c_1 T) \;,$$

*where the graph $G = (L, R, E)$ is sampled from $\mathbb{G}(n, n, M)$.*

**Proof of Lemma 4.4.** The distribution of the cuckoo graph given the fact $|E| = M'$ is identical to the distribution of $\mathbb{G}(n, n, M')$. Let us sample the graph $G_1$ from $\mathbb{G}(n, n, M')$ and the graph $G_2$ from $\mathbb{G}(n, n, M)$, such that $M' \leq M$ and assume that $G_1$ and $G_2$ were sampled such that they both satisfy the conditions of Corollary 6.4. Then

$$\Pr\left[\sum_{i=1}^{T} |C_{G_1}(v_i)| \geq c_2 T\right] \leq \Pr\left[\sum_{i=1}^{T} |C_{G_2}(u_i)| \geq c_2 T\right],$$

for $v_i \in V(G_1)$ and $u_i \in V(G_2)$ and under the conditions of Corollary 6.4. This is because the probability, that the sum of sizes of connected components is larger than a certain threshold, grows as we add edges to the graph in the course of the random graph process. Since in the cuckoo graph there are at most $(1 - \epsilon)n$ edges and $L = R = [n]$, Lemma 4.4 follows. ■

## 7 Experimental Results

In this section we demonstrate via experiments that our data structure is indeed very efficient, and can be used in practice. For simplicity, in our experiments we did not implement a cycle-detection mechanism, and used a fixed threshold instead: whenever the "age" of an element exceeded the threshold, we considered this element as a part of a second cycle in its connected component (this policy was suggested by Kirsch at el. [KMW08]). We note that this approach can only hurt the performance of our data structure, but our experimental results indicate that there is essentially no loss in forcing this simple policy. For the hash functions we used the keyed variant of the SHA-1 hash function, due to its good performance and freely available optimized versions. Figure 4 presents the parameters of our experiments.

| Parameter | Meaning |
|---|---|
| $n$ | The number of elements |
| $\epsilon$ | The stretch factor for each table |
| $MaxIter$ | The number of moves before stashing an element |
| $L$ | The number of iterations we perform per insert operation |
| $NumOfRuns$ | The number of times we repeat the experiment |

**Figure 4:** The parameters of our experiments.

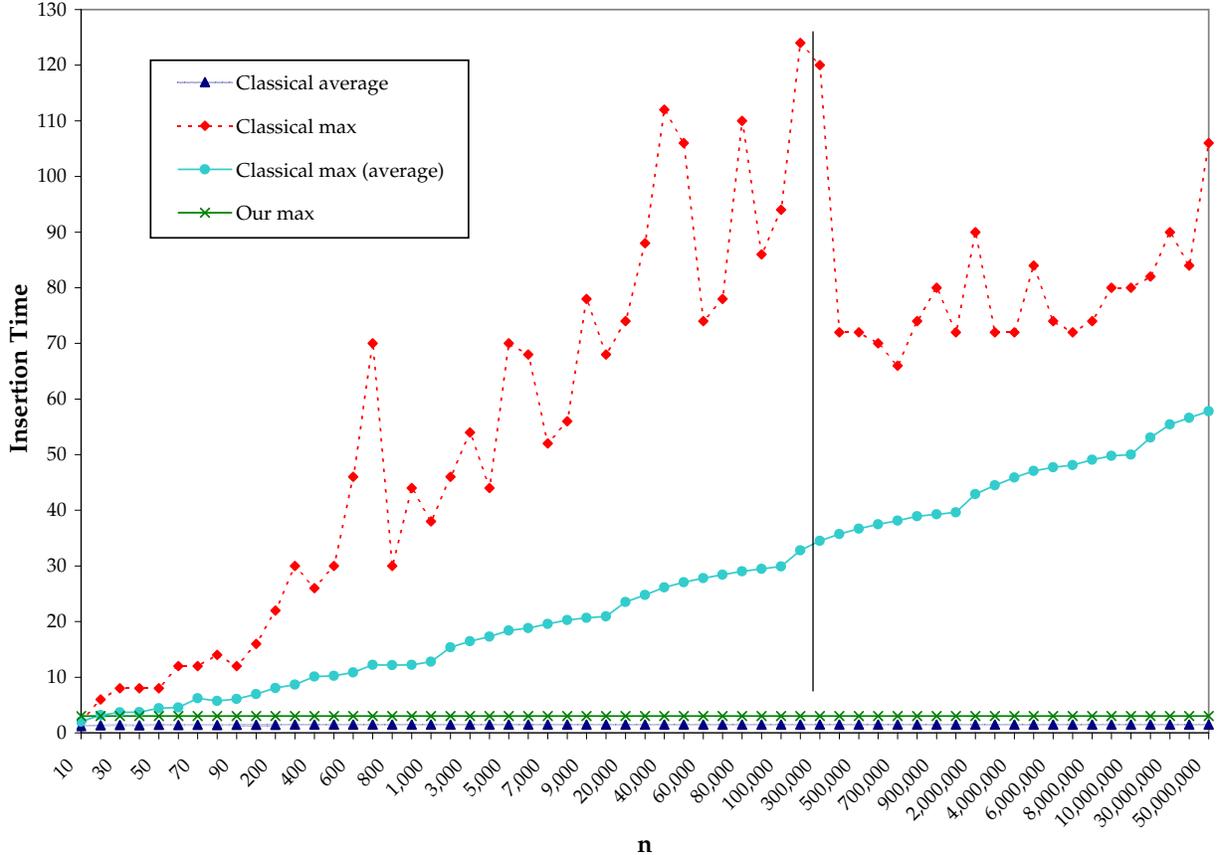

**Figure 5:** The insertion time of classical cuckoo hashing vs. our dictionary for $\epsilon = 0.2$.

Figure 5 presents a comparison between the insertion time of our scheme and the classical cuckoo hashing. In both cases we created a pseudorandom permutation $\sigma \in S_n$, and executed $n$ insert operations, where in step $i$ the element $\sigma(i)$ was inserted. The insertion time is measured as the length of the path that an element has traversed in the cuckoo graph. For the classical cuckoo hashing we show three curves: the average insertion time, the maximal insertion time and the average maximal insertion time. For our dictionary, the maximal insertion time is plotted, which was set to 3. We used $\epsilon = 0.2$, and $NumOfRuns$ was 100000 for $n = 10, \ldots, 300000$ and 1000 for $n = 400000, \ldots, 50000000$ (the vertical line indicates the transition point).

Figure 6 shows the size of the queue in our dictionary for $\epsilon = 0.2$. We show the average size, the maximal size and the average maximum. As before, the vertical line indicates the transition point in $NumOfRuns$. Note that the scale in the graphs is logarithmic (more precisely, log-lin scale), since we wanted to show the results for the whole range of $n$'s.

We observed a connection between the average maximal size of the queue in our dictionary and the average maximal insertion time in the classical cuckoo hashing: both behaved very close to $c \log_2 n$ for $c < 2.3$. In addition, in experiments which we don't present here graphically, we observed an expected tradeoff between $L$ and the maximal size of the queue (the queue grows larger as $L$ decreases).

Our experiments showed an excellent performance of our dictionary. After trying different values of $L$ we observed that a value as small as 3 is sufficient. This clearly demonstrates that adding an auxiliary memory of small (up to logarithmic) size reduces the worst case insertion time from logarithmic to a tiny constant.

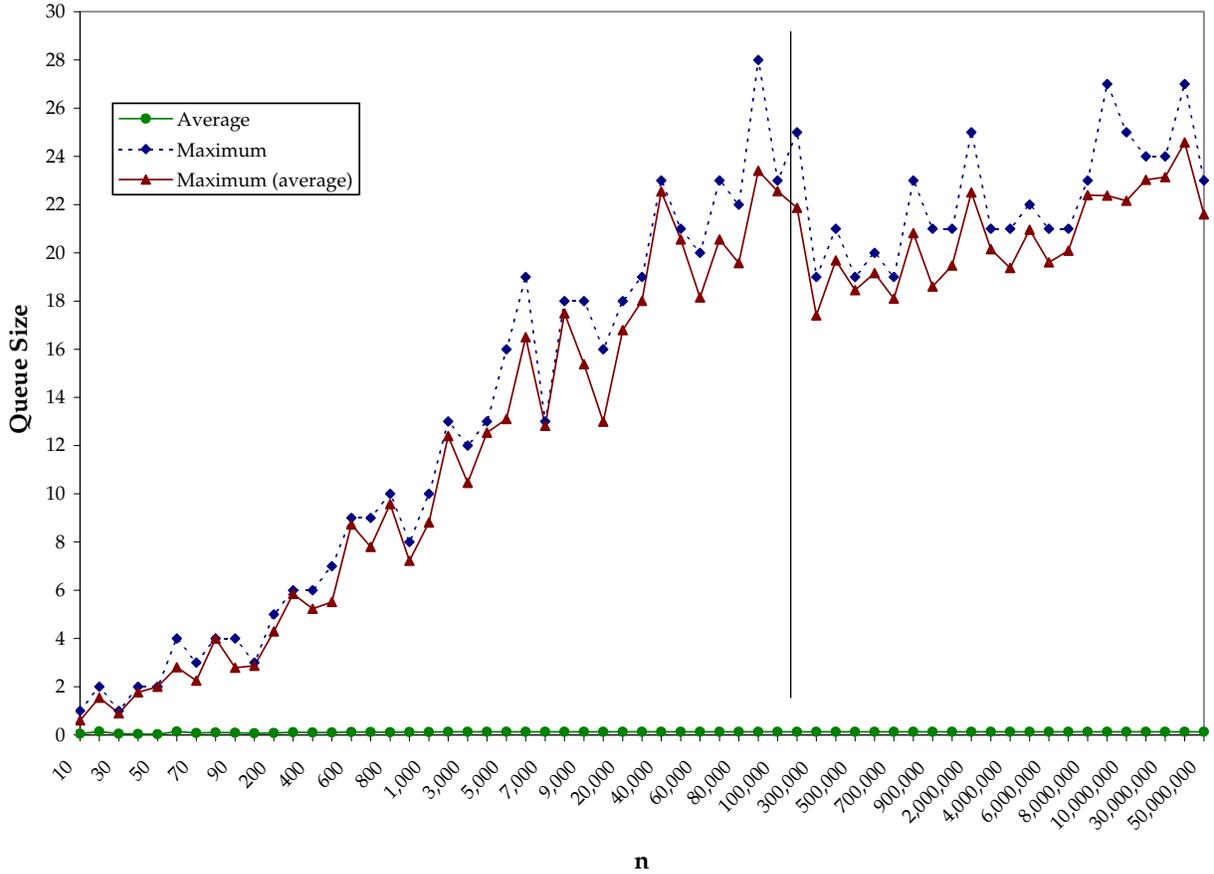

**Figure 6:** The size of the queue in our dictionary for $\epsilon = 0.2$.

## 8 Concluding Remarks

**Clocked adversaries.** The worst case guarantees of our dictionary are important if one wishes to protect against "clocked adversaries", as discussed in Section 1. In the traditional RAM model, such guarantees are also sufficient for protecting against such attacks. However, for modern computer architectures the RAM model has limited applicability, and is nowadays replaced by more accurate hierarchical models (see, for example, [AAC+87]), that capture the effect of several cache levels. Although our construction enables the "brute force" solution that measures the exact time every operation takes (see Section 1), a more elegant solution is desirable, which will make a better utilization of the cache hierarchy. We believe that our dictionary is an important step in this direction.

**Memory utilization.** Our construction achieves memory utilization of essentially 50%. More efficient variants of cuckoo hashing [FPS+05, Pan05, DW07] circumvent the 50% barrier and achieve better memory utilization by either using more than two hash functions, or storing more than one element in each entry. As demonstrated by Kirsch and Mitzenmacher [KM07], queue-based de-amortization performs very well in practice on these generalized variants, and it would be interesting to extend our analysis to these variants.

## Acknowledgments

We thank Michael Mitzenmacher, Eran Tromer and Udi Wieder for very helpful discussions concerning queues, stashes and caches.